\documentclass[twocolumn]{aastex631} %
\received{2021 August 11}
\revised{2021 September 9}
\accepted{2021 September 19}

\usepackage{graphicx}	% Including figure files
\usepackage{amsmath}	% Advanced maths commands
\usepackage[normalem]{ulem}
\usepackage{booktabs}	
\usepackage{mathtools}	
\usepackage{color}
\usepackage{acronym}   
\usepackage{xspace}
\newcommand{\monei}{\ensuremath{m_{1,\rm{ZAMS}}}\xspace}
\newcommand{\mtwoi}{\ensuremath{m_{2,\rm{ZAMS}}}\xspace}
\newcommand{\monef}{\ensuremath{m_{1,\rm{f}}}\xspace}
\newcommand{\mtwof}{\ensuremath{m_{2,\rm{f}}}\xspace}
\newcommand{\ai}{\ensuremath{a_{\rm{ZAMS}}}\xspace}
\newcommand{\qi}{\ensuremath{q_{\rm{ZAMS}}}\xspace}
\newcommand{\Zi}{\ensuremath{Z}\xspace}

\newcommand{\km}{\ensuremath{\,\rm{km}}\xspace}
\newcommand{\kms}{\ensuremath{\,\rm{km}\,\rm{s}^{-1}}\xspace}
\newcommand{\Msun}{\ensuremath{\,\rm{M}_{\odot}}\xspace}

\newcommand{\Zsun}{\ensuremath{\,\rm{Z}_{\odot}}\xspace}

\newcommand{\AU}{\ensuremath{\,\mathrm{AU}}\xspace}

\newcommand{\SFRD}{\text{SFRD}\ensuremath{(Z,z)}\xspace}

\newcommand{\PII}{Broekgaarden et al. (in prep.)}
\newcommand{\PIItep}{(Broekgaarden et al. 2021, in preperation)}

\newcommand{\mnsf}{\ensuremath{m_{\rm{NS}}}\xspace}

\newcommand{\mbhf}{\ensuremath{m_{\rm{BH}}}\xspace}

\newcommand{\mtotf}{\ensuremath{m_{\rm{tot}}}\xspace}
\newcommand{\mchirpf}{\ensuremath{{m}_{\rm{chirp}}}\xspace}

\newcommand{\qf}{\ensuremath{q_{\rm{f}}}\xspace}

\newcommand{\chibh}{\ensuremath{{\chi}_{\rm{1}}}\xspace}
\newcommand{\Rns}{\ensuremath{{R}_{\rm{NS}}}\xspace}
\newcommand{\Rgwone}{\ensuremath{\mathcal{R}_{\rm{GW200115}}}\xspace}
\newcommand{\Rgwzero}{\ensuremath{\mathcal{R}_{\rm{GW200105}}}\xspace}
\newcommand{\Rbhns}{\ensuremath{\mathcal{R}_{\rm{BHNS}}}\xspace}
\newcommand{\Rbhbh}{\ensuremath{\mathcal{R}_{\rm{BHBH}}}\xspace}
\newcommand{\Rnsns}{\ensuremath{\mathcal{R}_{\rm{NSNS}}}\xspace}

\newcommand{\Nmodels}{\ensuremath{560}\xspace}
\newcommand{\NmodelsBPS}{\ensuremath{20}\xspace}
\newcommand{\NmodelsMSSFR}{\ensuremath{28}\xspace}
\newcommand{\gwone}{\ensuremath{\rm{GW200115}}\xspace} 
\newcommand{\gwzero}{\ensuremath{\rm{GW200105}}\xspace} 
\newcommand{\Gpcyr}{\ensuremath{\,\rm{Gpc}^{-3}\,\rm{yr}^{-1}}\xspace}
\newcommand{\model}{P112\xspace}

\acrodef{GSMF}{galaxy stellar mass function, the number density of galaxies per logarithmic mass bin,}
\acrodef{MZR}{mass-metallicity relation}
\acrodef{SFRD}{star formation rate density}
\acrodef{BHNS}{black hole--neutron star}
\acrodef{NSNS}{binary neutron star}
\acrodef{BHBH}{binary black hole}
\acrodef{DCO}{double compact object}
\acrodef{NS}{neutron star}
\acrodef{BH}{black hole}
\acrodef{BH--NS}{black hole-neutron star}
\acrodef{GRB}{gamma-ray burst}
\acrodef{RLOF}{Roche-lobe overflow}
\acrodef{CE}{common envelope}
\acrodef{GW}{gravitational-wave}
\acrodefplural{GW}[GWs]{gravitational waves}
\acrodef{SN}{supernova}
\acrodefplural{SN}[SNe]{supernovae}
\acrodef{ECSN}{electron-capture SN}
\acrodef{PISN}{pair-instability SN}
\acrodefplural{ECSN}[ECSNe]{electron-capture SN}
\acrodef{USSN}{ultra-stripped SN}
\acrodefplural{USSN}[USSNe]{ultra-stripped SN}
\acrodef{CCSN}{core-collapse SN}
\acrodefplural{CCSN}[CCSNe]{core-collapse SN}
\acrodef{COMPAS}{
Compact Object Mergers: Population Astrophysics and Statistics}
\acrodef{SFRD}{metallicity-specific star formation rate density}
\acrodef{ZAMS}{zero-age main sequence}

\shorttitle{Formation of GW200115 and GW200105}
\shortauthors{Broekgaarden $\&$ Berger}

\begin{document}

% Title of the paper, and the short title which is used in the headers.
% Keep the title short and informative.
\title{Formation of the First Two Black Hole--Neutron Star Mergers (GW200115 and GW200105) from Isolated Binary Evolution}

% The list of authors, and the short list which is used in the headers.
% If you need two or more lines of authors, add an extra line using \newauthor
\author[0000-0002-4421-4962]{Floor S. Broekgaarden}
\affiliation{Center for Astrophysics \textbar{} Harvard $\&$ Smithsonian, 60 Garden Street, Cambridge, MA 02138, USA; \url{floor.broekgaarden@cfa.harvard.edu}}

\author[0000-0002-9392-9681]{Edo Berger}
\affiliation{Center for Astrophysics \textbar{} Harvard $\&$ Smithsonian, 60 Garden Street, Cambridge, MA 02138, USA; \url{floor.broekgaarden@cfa.harvard.edu}}

\begin{abstract}
In this work we study the formation of the first two \ac{BHNS} mergers detected in gravitational waves (\gwone and \gwzero) from massive stars in wide isolated binary systems -- the \textit{isolated binary evolution channel}. We use \Nmodels \ac{BHNS} binary population synthesis model realizations and show that the system properties (chirp mass, component masses, and mass ratios) of both \gwone and \gwzero match predictions from the isolated binary evolution channel. We also show that most model realizations can account for the local \ac{BHNS} merger rate densities inferred by LIGO--Virgo. However, to simultaneously also match the inferred local merger rate densities for BHBH and NSNS systems we find we need models with moderate kick velocities ($\sigma\lesssim 10^2$\kms) or high common-envelope efficiencies ($\alpha_{\rm{CE}}\gtrsim 2$) within our model explorations.  We conclude that the first two observed \ac{BHNS} mergers can be explained from the isolated binary evolution channel for reasonable model realizations.
\end{abstract}

%%%%%%%%%%%%%%%%%%%%%%%%%%%%%%%%%%%%%%%%%%%%%%%%%%

%%%%%%%%%%%%%%%%% BODY OF PAPER %%%%%%%%%%%%%%%%%%

%%%%%%%%%%%%%%%%%%%%%%%%%%%%%%%%%%%%%%%%%%%%%
%%%%%%%%%%%%%%%%%%%%%%%%%%%%%%%%%%%%%%%%%%%%%
%%%%%%%%%%%%%%% INTRODUCTION %%%%%%%%%%%%%%%%
%%%%%%%%%%%%%%%%%%%%%%%%%%%%%%%%%%%%%%%%%%%%%
%%%%%%%%%%%%%%%%%%%%%%%%%%%%%%%%%%%%%%%%%%%%%

%@arxiver{Scatter_Final_kde_with_LV_112_P.png,Rates_intrinsic_single_panel.pdf}

\section{Introduction}
\label{sec:introduction}
In 2021 June, \citet{Abbott:2021-first-NSBH} announced the first observations of \acp{GW} from  two \ac{BHNS} merger events--\gwone and \gwzero--during the third LIGO--Virgo--Kagra (LVK) observing run (O3). \gwone was detected by all three detectors from LIGO and Virgo and has chirp mass\footnote{Throughout this paper we use the reported `high spin' source parameters. This assumption does not significantly impact our results. The values reported are the median and $90\%$ credible intervals.} $\mchirpf = 2.42_{-0.07}^{+0.05}$\Msun, total mass $\mtotf = 7.1_{-1.4}^{+1.5}$\Msun, component masses $\mbhf = 5.{7}_{-2.1}^{+1.8}$ $\Msun$ and $\mnsf = 1.{5}_{-0.3}^{+0.7}$\Msun and mass ratio $\qf \equiv (\mnsf / \mbhf) = 0.26_{-0.10}^{+0.35}$.  \gwzero was effectively only observed by LIGO--Livingston as LIGO--Hanford was offline and the signal-to-noise ratio in Virgo was below the threshold of 4.0. \gwzero has $\mchirpf = 3.41^{+0.08}_{-0.07}$, $\mtotf = 10.9_{-1.2}^{+1.1}$\Msun, $\mbhf = 8.{9}_{-1.5}^{+1.2}$ $\Msun$, $\mnsf = 1.{9}_{-0.2}^{+0.3}$ $\Msun$ and $\qf = 0.22_{-0.04}^{+0.08}$. 

From these observations \citet{Abbott:2021-first-NSBH} infer a local \ac{BHNS} merger rate density of $\Rbhns = {45}_{-33}^{+75}$\Gpcyr when assuming that \gwone and \gwzero are solely representative of the entire \ac{BHNS} population; and  $\Rbhns = {130}_{-69}^{+112}$\Gpcyr when assuming a broader distribution of component masses \citep{Abbott:2021-first-NSBH}. For the individual \gwone and \gwzero events, the authors quote inferred local merger rate densities of  $\Rgwone = {36}_{-30}^{+82}$\Gpcyr and $\Rgwzero = {16}_{-14}^{+38}$\Gpcyr, respectively. In \citet{GWTC2:pop}, LVK reported a \ac{NSNS} merger rate of $\Rnsns = 320^{+490}_{-240}$\Gpcyr, and four different \ac{BHBH} $90\%$ credible rate intervals spanning $\Rbhbh\approx 10.3-104$\Gpcyr.

The main formation channel leading to merging \ac{BHNS} systems (and BHBH and NSNS) is still under debate. A widely studied channel is the formation of \ac{BHNS} mergers from massive stars that form in (wide) isolated binaries and evolve typically including a \ac{CE} phase  \citep[e.g.][]{Neijssel:2019,Belczynski:2020,Shao:2021}. Other possible channels include formation from  close binaries that can evolve chemically homogeneously \citep{MandelDeMink:2016,Marchant:2017}, metal-poor Population III stars that formed in the early universe \citep[e.g.][]{Belczynski:2017popIII}, stellar triples or multiples \citep[][]{FragioneLoeb:2019a,HamersThompson:2019,Hamers:2021}, or from dynamical or hierarchical interactions in globular clusters \citep[][]{Clausen:2013, ArcaSedda:2020, Ye:2019}, nuclear star clusters \citep[][]{PetrovichAntonini:2017, McKernan:2020, Wang:2020} and young and/or open star clusters \citep[e.g.,][]{Ziosi:2014,Rastello:2020,ArcaSedda:2021}. We refer the reader to \citet[][]{MandelBroekgaardenReview:2021} for a living review of these various formation channels.  

In this Letter we address the key question: {\emph{ Could GW200115 and GW200105 have been formed through the isolated binary evolution scenario?}} 

To investigate this we use the simulations from \citet{ZenodoDCOBHNS:2021} to study the formation of merging \ac{BHNS} systems from pairs of massive stars that evolve through the isolated binary evolution scenario. This Letter is structured as follows. In \S\ref{sec:method} we describe our method and models.  In \S\ref{sec:results-intrinsic-merger-rates} we show that most of our models do match the inferred \ac{BHNS} rate densities, but that only models with higher \ac{CE} efficiencies or moderate \ac{SN} kicks are also consistent with the inferred \Rbhbh and \Rnsns. In \S\ref{sec:results-matching-the-GW-properties} we compare the properties of \gwone and \gwzero to the overall expected \ac{GW}-detectable \ac{BHNS} population.  We end with a discussion in \S\ref{sec:discussion} and present our conclusions in~\S\ref{sec:conclusions}.

%%%%%%%%%%%%%%%%%%%%%%%%%%%%%%%%%%%%%%%%%%%%%
%%%%%%%%%%%%%%%%%%%%%%%%%%%%%%%%%%%%%%%%%%%%%
%%%%%%%%%%%%%%%     METHOD   %%%%%%%%%%%%%%%%
%%%%%%%%%%%%%%%%%%%%%%%%%%%%%%%%%%%%%%%%%%%%%
%%%%%%%%%%%%%%%%%%%%%%%%%%%%%%%%%%%%%%%%%%%%%

\section{Method}
\label{sec:method}

We use the publicly available binary population synthesis simulations from \citet[][presented in  Broekgaarden et al., in preparation]{ZenodoDCOBHNS:2021}, to study the formation of \gwone and \gwzero from the isolated binary evolution channel.  The simulations used in this work 
add new model realizations compared to \citet{Broekgaarden:2021}, and also consider merging \ac{BHBH} and \ac{NSNS} systems. The simulations are performed using the rapid binary population synthesis code {\sc{COMPAS}}  \citep[][]{Stevenson:2017, Barrett:2017, VignaGomez:2018, Broekgaarden:2019, Neijssel:2019}, which is used to model the evolution of the binary systems and determine the source properties and rates of the double compact object mergers. The \ac{BHNS} population data set contains a total of \Nmodels  model realizations to explore the uncertainty in the population modeling. 
Namely, \NmodelsBPS different binary population synthesis variations (varying assumptions for common envelope, mass transfer, supernovae, and stellar winds) and \NmodelsMSSFR model variations in the metallicity-specific star formation rate density model, \SFRD (varying assumptions for the star formation rate density, mass-metallicity relation, and galaxy stellar mass function), which is a function of birth metallicity ($Z$) and redshift ($z$). The population synthesis simulations are labeled A, B, C, ... T, with each variation representing one change in the physics prescription compared to the fiducial model `A' (see Table~1 in Broekgaarden et al. in preparation); the \SFRD models are labelled with 000, 111, 112, ... 333 (see Table~3 \citealt[][]{Broekgaarden:2021}).  To obtain high--resolution simulations, \PII{} simulated for each population synthesis model a million binaries for 53 $\Zi$ bins and used the adaptive importance sampling algorithm STROOPWAFEL \citep{Broekgaarden:2019} to further increase the number of \ac{BHNS} systems in the simulations. Doing so, resulted in a total dataset consisting of over 30 million \ac{BHNS} systems, making it the most extensive simulation of its kind to date. 

We define \ac{BHNS} systems in our simulations to match the observed \gwone and \gwzero if their \mchirpf, \mtotf, \monef, \mtwof \'{a}nd \qf lie within the inferred $90\%$ credible intervals (\S\ref{sec:introduction}). We note that \citet{Abbott:2021-first-NSBH} also inferred $90\%$ credible intervals for the spins of both \ac{BHNS} systems, but due to the large uncertainties in the measurements and the theory of spins we leave this topic for discussion in \S\ref{sec:discussion} and do not explicitly take spins into account for the \ac{BHNS} system selection. We calculate \Rbhns using Equation~2 in  \citet{Broekgaarden:2021}, where we assume a local redshift $z\approx 0$, and discuss these intrinsic merger rates in \S\ref{sec:results-intrinsic-merger-rates}. We obtain the detection-weighted distributions for the \ac{BHNS} mergers using Equation~3 from \citet{Broekgaarden:2021} and discuss these in \S\ref{sec:results-matching-the-GW-properties}. To calculate the detectable \ac{GW} population we assume the sensitivity of a GW--detector network equivalent to advanced LIGO in its design configuration \citep{2015CQGra..32g4001L, 2016LRR....19....1A, 2018LRR....21....3A}, a reasonable proxy for O3. For the purpose of comparison, we use the LIGO--Virgo posterior samples for \gwone and \gwzero from \citet{Abbott:2021-open-GWTC-data}.

\section{Predicted BHNS Merger Rates and Properties}
\label{sec:results}

\begin{figure*}
    \centering
\includegraphics[width=1.0\textwidth]{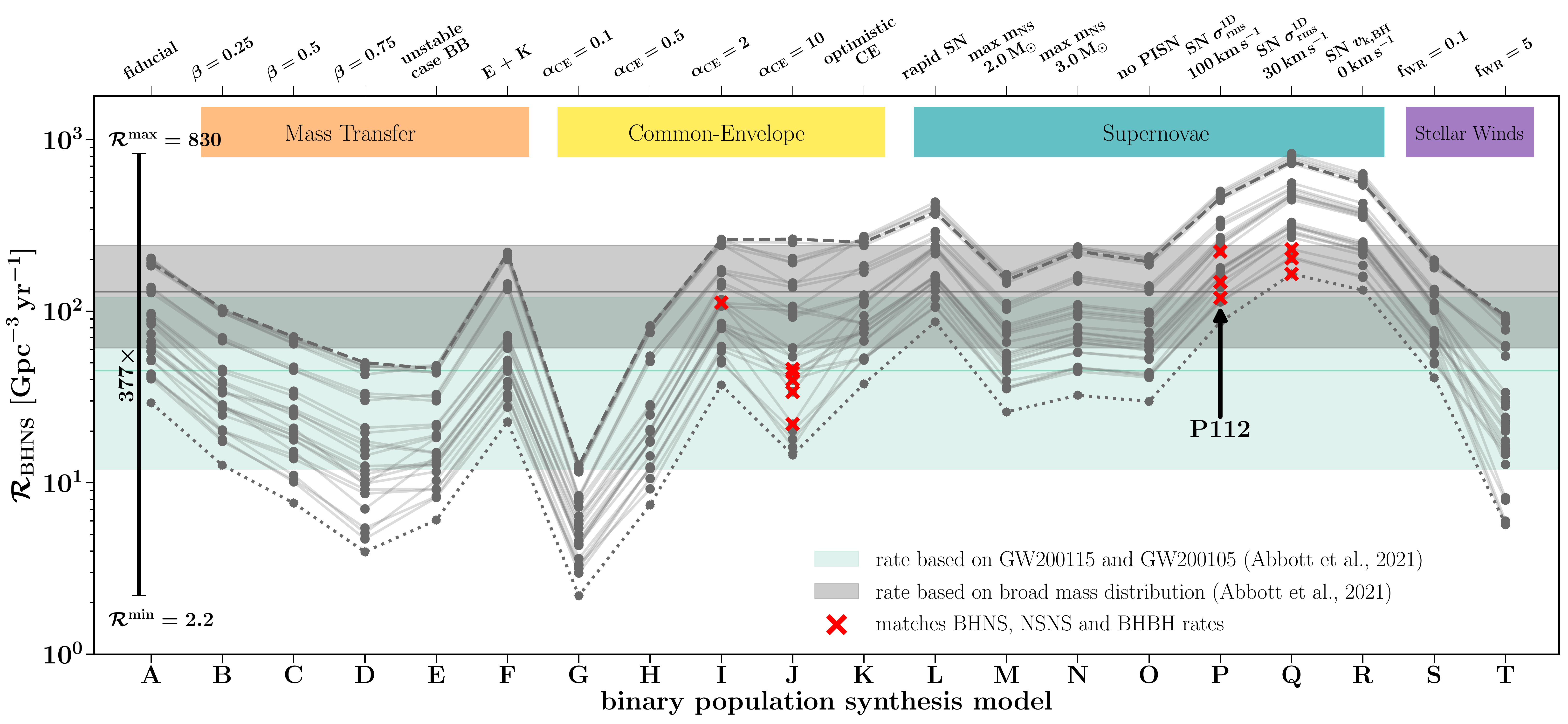} 
    \caption{Predicted local \ac{BHNS} merger rate density, \Rbhns, for our \Nmodels model variations.
    The shaded horizontal bars mark the corresponding \ac{GW}-inferred $90\%$ credible intervals for the merger rate densities from \citet[][]{Abbott:2021-first-NSBH}: $\Rbhns = {45}_{-33}^{+75}$\Gpcyr (green) and  $\Rbhns = {130}_{-69}^{+112}$\Gpcyr (gray). 
    We connect simulation predictions that use the same \SFRD model with a line for visual clarity only. Two \SFRD variations, 231 (dashed) and 312 (dotted),  are highlighted.
    Model realizations matching one of the inferred \Rbhns and also the $90\%$ credible intervals for the \ac{BHBH} and \ac{NSNS} merger rates from \citet[][]{GWTC2:pop} are marked with red crosses.
    In the top we added colored labels to indicate what physics assumptions are varied compared to our fiducial assumptions in the models. An arrow points to model \model (\S\ref{sec:results-matching-the-GW-properties}).
    }
    \label{fig:Rates-Intrinsic}
\end{figure*}

\subsection{Local BHNS merger rates}
\label{sec:results-intrinsic-merger-rates}

In Figure~\ref{fig:Rates-Intrinsic} we show the predicted local merger rate densities from our \Nmodels model realizations for the overall \ac{BHNS} population, in comparison to the $90\%$ credible intervals from  \citet{Abbott:2021-first-NSBH}. We find that the majority of the \Nmodels model realizations match one of the two observed \ac{BHNS} merger rate densities. Model realizations that underpredict the observed rates include most \SFRD variations of model G ($\alpha_{\rm{CE}}=0.1$) corresponding to inefficient \ac{CE} ejection, which increases the number of stellar mergers during the \ac{CE} phase (our fiducial model uses $\alpha_{\rm{CE}}=1$), and about half of the \SFRD variations of model D, which assumes a high mass transfer efficiency ($\beta = 0.75$), as opposed to our fiducial model that assumes an adaptive $\beta$ based on the stellar type and thermal timescale and typically results in $\beta\lesssim 0.1$ for systems leading to \ac{BHNS} mergers.  Conversely, some model realizations overpredict the observed rates, in particular about half of the \SFRD variations of models P, Q and R.  These models have moderate or low \ac{SN} natal kick magnitudes, increasing the number of \ac{BHNS} systems that stay bound during the \acp{SN}. The \SFRD variations that overpredict the observed rates correspond to lower average metallicities, thereby increasing the formation efficiency of \ac{BHNS} mergers \citep{Broekgaarden:2021}.

On the other hand, we find that only a small subset of the \Nmodels model realizations (shown with red crosses in Figure~\ref{fig:Rates-Intrinsic})  also match the inferred $90\%$ credible intervals of the observed \ac{BHBH} and \ac{NSNS} merger rate densities (\S\ref{sec:introduction})\footnote{We note that \Rbhbh, \Rbhns and \Rnsns could have (large) contributions from formation channels other than the isolated binary evolution channel.},
namely, models I, J,  P and Q in conjunction with a few of the \SFRD variations. Both the higher $\alpha_{\rm{CE}}$ values in models I and J ($\alpha_{\rm{CE}}\gtrsim 2$), and the low \ac{SN} natal kicks in models P and Q ($\sigma\approx 30$ or $100\kms$, where $\sigma$ is the one-dimensional rms velocity dispersion of the Maxwellian distribution used to draw the \ac{SN} natal kick magnitudes), result in relatively higher \ac{NSNS} rates that can match the high observed \Rnsns\footnote{Most isolated binary evolution predictions (including most of our model variations) underestimate the inferred \ac{NSNS} merger rate (e.g., \citealt{Chruslinska:2018,MandelBroekgaardenReview:2021}).}$^{,}$\footnote{We note that there have been several recent studies supporting common-envelope efficiencies $\alpha_{\rm{CE}} \gtrsim 2$ \citep[e.g.][]{Fragos:2019,Garcia:2021,Schreier:2021}.}.  
Requiring a match with the observed \Rbhbh mostly constrains the \SFRD models to those with moderate average star formation metallicities, as our models with typically lower \Zi\footnote{E.g., the models that assume a galaxy mass-metallicity relation based on \citet[][]{2006ApJ...638L..63L} (all \SFRD models $\rm{xyz}$ with $\rm{z}=1$), which maps to lower average stellar birth metallicities (for example, model 231  has an average star formation \Zi of $\approx \Zsun/10$ near redshift $z \approx 2$, whereas for our higher \Zi models this is closer to $\Zi \approx \Zsun$ around the same redshift).  }
overestimate the inferred \Rbhbh. Similar results were found by earlier work including \citet{GiacobboMapelli:2018} and \citet{Santoliquido:2021-isolated-binaries}.  

\begin{figure*}
    \centering
\includegraphics[width=.75\textwidth]{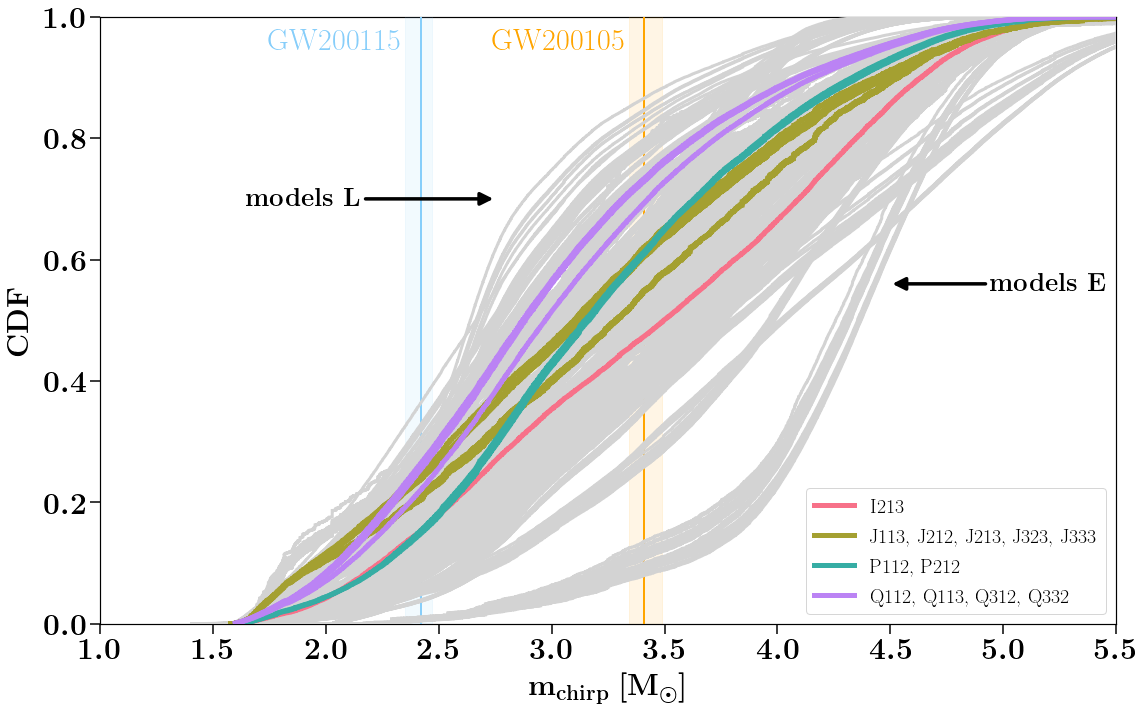}
    \caption{Cumulative distributions of the chirp mass for the models matching \Rbhns, \Rbhbh and \Rnsns (colored lines) and all of the other \Nmodels model realizations (light gray lines). We also show the $90\%$ credible intervals for \gwone and \gwzero (vertical bars; \citealt{Abbott:2021-first-NSBH}). The legend indicates the label names of the matching models, while the arrows point to models E and L, which predict the lowest and highest fraction of \ac{BHNS} mergers within the chirp mass range spanning \gwone and \gwzero, respectively.
    }
    \label{fig:chirp-mass-cdf-matching-models}
\end{figure*}

Within the matching models, models I, P and Q match the inferred \Rbhns that is based on a broader \ac{BHNS} mass distribution, whereas the matching model J variations overlap only with the observed rate based on a \gwone- and \gwzero-like population. We note, however, that our binary population synthesis models in all cases predict a broader mass distribution compared to just \gwone- and \gwzero-like events.  We investigate this in detail in Figure~\ref{fig:chirp-mass-cdf-matching-models}, where we plot the cumulative \ac{BHNS} chirp mass distributions of our model variations, in comparison to the chirp masses spanned by \gwone and \gwzero, $2.35\lesssim \mchirpf / \Msun \lesssim 3.49$.  We find that $\approx 60\%$ of the \ac{GW}-detectable \ac{BHNS} systems in model J are expected to have \mchirpf outside of this range, while for matching models I, P and Q this is about $60\%$, $50\%$, and $50\%$, respectively. For models I, P and Q this result is expected since they match the \Rbhns range that is based on a broader mass distribution, but for model J the low percentage of $60\%$ conflicts the match with \Rbhns based on a \ac{BHNS} population defined by \gwone- and \gwzero-like events. From Figure~\ref{fig:chirp-mass-cdf-matching-models} it can be seen that besides models I, J, P and Q all other model realizations generally predict \ac{BHNS} populations with broader chirp mass distributions compared to the range spanned by \gwone and \gwzero alone.  The models using the rapid supernova prescription (model L) predict the highest fraction ($\approx 75\%$) of \ac{BHNS} systems with $2.35\lesssim \mchirpf / \Msun \lesssim 3.49$, whereas the model assuming that case BB mass transfer is always unstable (model E) results in the lowest percentages ($\approx 8\%$).

%%%%%%%%%%%%%%%%%%%%%%%%%%%%%%%%%%%%%%%%%%%%%%
%%%%%%        at MERGER                 %%%%%%
%%%%%%%%%%%%%%%%%%%%%%%%%%%%%%%%%%%%%%%%%%%%%%

\subsection{Properties of the BHNS systems}
\label{sec:results-matching-the-GW-properties}
In the following discussion we focus on the specific model `P112', as an example of a model realization that matches all of the various observed merger rate densities.  We take this approach for simplicity, but note that we are not claiming that only this model realization represents the correct isolated binary evolution pathway to the observed \ac{GW} mergers. 
Below we examine the properties of the systems at the time of merger (chirp mass, component masses and mass ratio), as well as at the time of formation on the \ac{ZAMS} (component masses, mass ratio and semimajor axis).

\subsubsection{BHNS properties at merger}
\label{results:BHNS-properties-at-merger}

\begin{figure*}
    \centering
\includegraphics[width=1.0\textwidth]{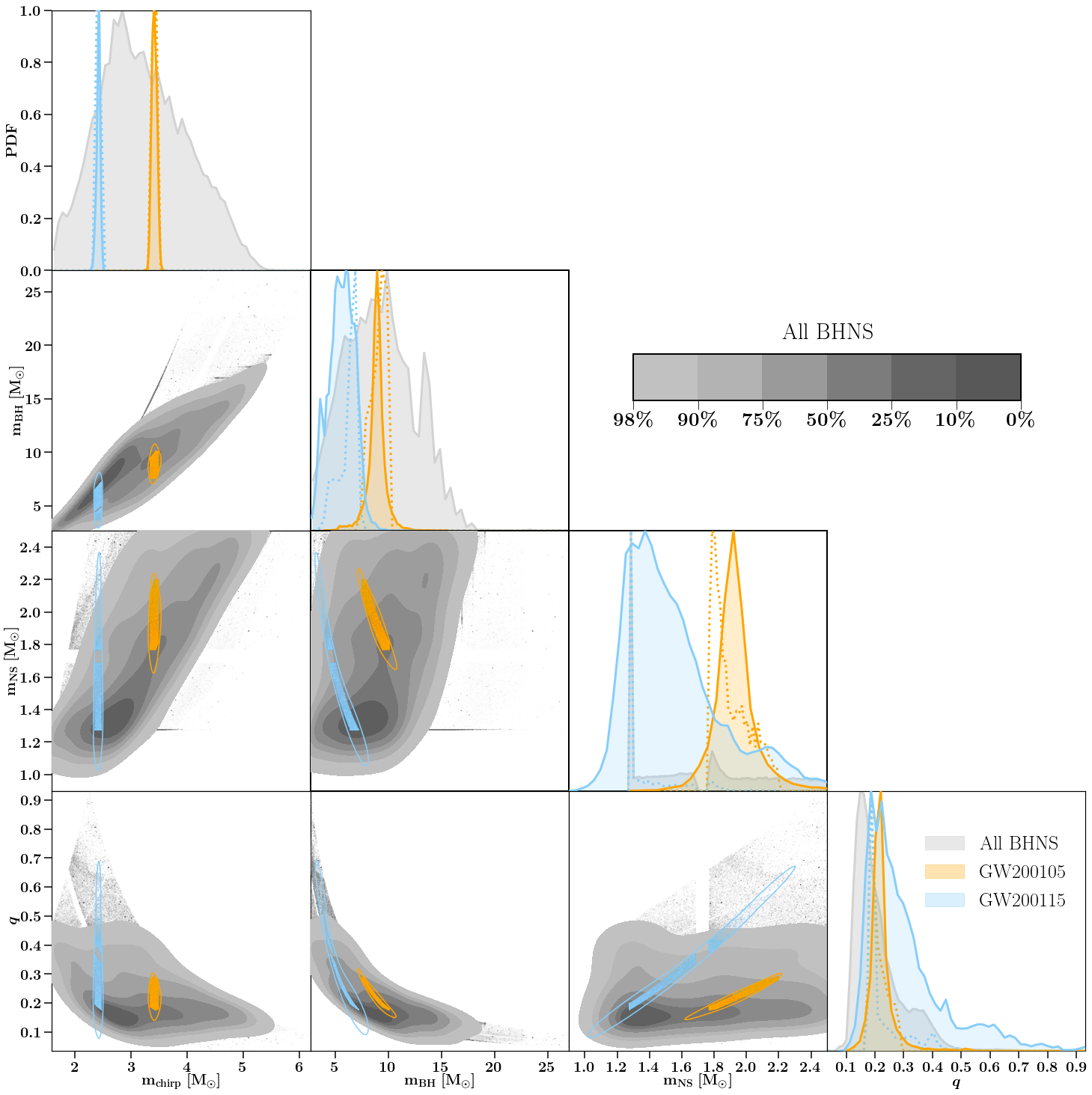} 
    \caption{Corner plot showing the 1D and 2D distributions of the properties of the detectable \ac{BHNS} mergers from our binary population synthesis model \model.  We show the chirp mass, \ac{BH} mass, \ac{NS} mass, and the mass ratio at the time of merger.  In gray we show the overall \ac{BHNS} population, whereas in blue (orange) we show \ac{BHNS} systems that have properties matching \gwone (\gwzero). Our \gwone (\gwzero) predictions are shown with blue (orange) scatter points and dotted histograms, whereas the posterior samples from \citet{Abbott:2021-first-NSBH} are shown with  $90\%$ contour levels in the 2D plots and with filled histograms in the 1D panels.  The gray contours show the percentage of the detectable \ac{BHNS} systems enclosed.  All distributions are weighted using the \ac{GW}-detection probability. The 1D distributions are normalized such that the peak is equal to one. 
    }
    \label{fig:Triangle-final}
\end{figure*}

In Figure~\ref{fig:Triangle-final} we show the 1D and 2D distributions of the predicted properties for the \ac{GW}-detectable \ac{BHNS} population for all \ac{BHNS} systems (gray contours and 1D distributions) and for \gwone- and \gwzero-like \ac{BHNS} systems (blue and orange scatter points and dotted histograms, respectively). The LIGO-Virgo inferred posterior samples for \gwone and \gwzero are shown with orange and blue $90\%$ credible contours in the 2D histograms and with filled histograms in the 1D plots, respectively. We show \mchirpf, \mbhf, \mnsf and \qf. In the top panels we normalize each 1D distribution to peak at a value of 1. 

Overall, we find that model \model predicts the majority ($90\%$ percentiles) of the \ac{GW}-detectable \ac{BHNS} mergers to have $2 \lesssim \mchirpf /\Msun \lesssim 4.6$,  $4.1 \lesssim \mbhf / \Msun \lesssim 14.7$,  $1.3\lesssim \mnsf / \Msun \lesssim 2.4$, and $0.1\lesssim\qf\lesssim 0.4$. We emphasize that the neutron star mass and lower black hole mass boundaries of $1$\Msun and $2.5$\Msun, respectively, are set by our binary population synthesis assumptions for the lower and upper \ac{NS} mass from the delayed \citet{Fryer:2012} remnant mass prescription. 

In detail, we find several interesting features in the model distributions compared to the observed \ac{BHNS} mergers. First, we note that the inferred properties of \gwone and \gwzero lie well within the predicted population of the \ac{GW}-detectable \ac{BHNS} population. In particular, the \gwone and \gwzero credible intervals typically overlap with the highest probability region for the corresponding distribution of the predicted \ac{BHNS} population. 
We stress that this result does not follow trivially from the match of model \model with the inferred \Rbhns (\S\ref{sec:results-intrinsic-merger-rates}) as the properties of the intrinsic and detectable \ac{BHNS} populations \emph{could} be significantly different due to the strong bias in the sensitivity of \ac{GW} detectors for more massive systems, meaning that the underlying intrinsic mass distributions can be significantly different from the observed mass distributions. Only for \mnsf the posterior samples of \gwone reach well below the predicted distribution of our models, but this is due to the remnant mass prescription, which has an artificial lower \mnsf limit of about $1.3$\Msun. The overlap between our predictions and the inferred posterior distributions can also be seen from the matches between the LVK distributions and our model-weighted distributions for \gwone and \gwzero. 

Second, we find that model \model suggests the existence of a small, positive, $\mbhf$--$\mnsf$ correlation in the GW-detectable \ac{BHNS} population (a similar correlation is also visible in the $\mchirpf$--$\mnsf$ distribution, but we note that the chirp mass is dependent on \mnsf). This means that we expect, on average, that \ac{BHNS} with more massive \acp{BH} have more massive \acp{NS}. Interestingly, this correlations also holds for \gwone and \gwzero. This correlation is visible in most of our other model variations, and was also noted by earlier work, including \citet{Kruckow:2018} and \citet{Broekgaarden:2021}. The correlation is due to the preference in the isolated binary evolution channel for more equal mass binaries. The \ac{BHNS} with more massive \ac{BH} typically form from binaries with a more massive primary (the initially more massive star), and such systems also have on average more massive secondaries at ZAMS (see \citealt{Sana:2012}). In addition, the more massive secondaries at ZAMS typically lead to binaries with more equal mass ratios at the moment of the first mass transfer, making it likely more stable and successfully leading to a \ac{BHNS}  \citep{Broekgaarden:2021}. This results on average in a more massive \ac{NS} in binaries with a more massive \ac{BH}.

Finally, we note that several of the panels in Figure~\ref{fig:Triangle-final} show sharp gaps or peaks in the distributions, particularly visible in the scatter points and 1D histograms. These gaps are artificial discontinuities present in some of the prescriptions in our COMPAS model \citep[see][and references therein]{Broekgaarden:2021}.

%%%%%%%%%%%%%%%%%%%%%%%%%%%%%%%%%%%%%%%%%%%%%%
%%%%%%        at ZAMS                    %%%%%
%%%%%%%%%%%%%%%%%%%%%%%%%%%%%%%%%%%%%%%%%%%%%%
\subsubsection{BHNS properties at ZAMS }
\label{results:BHNS-properties-at-ZAMS}

\begin{figure*}
    \centering
    \includegraphics[width=1.0\textwidth]{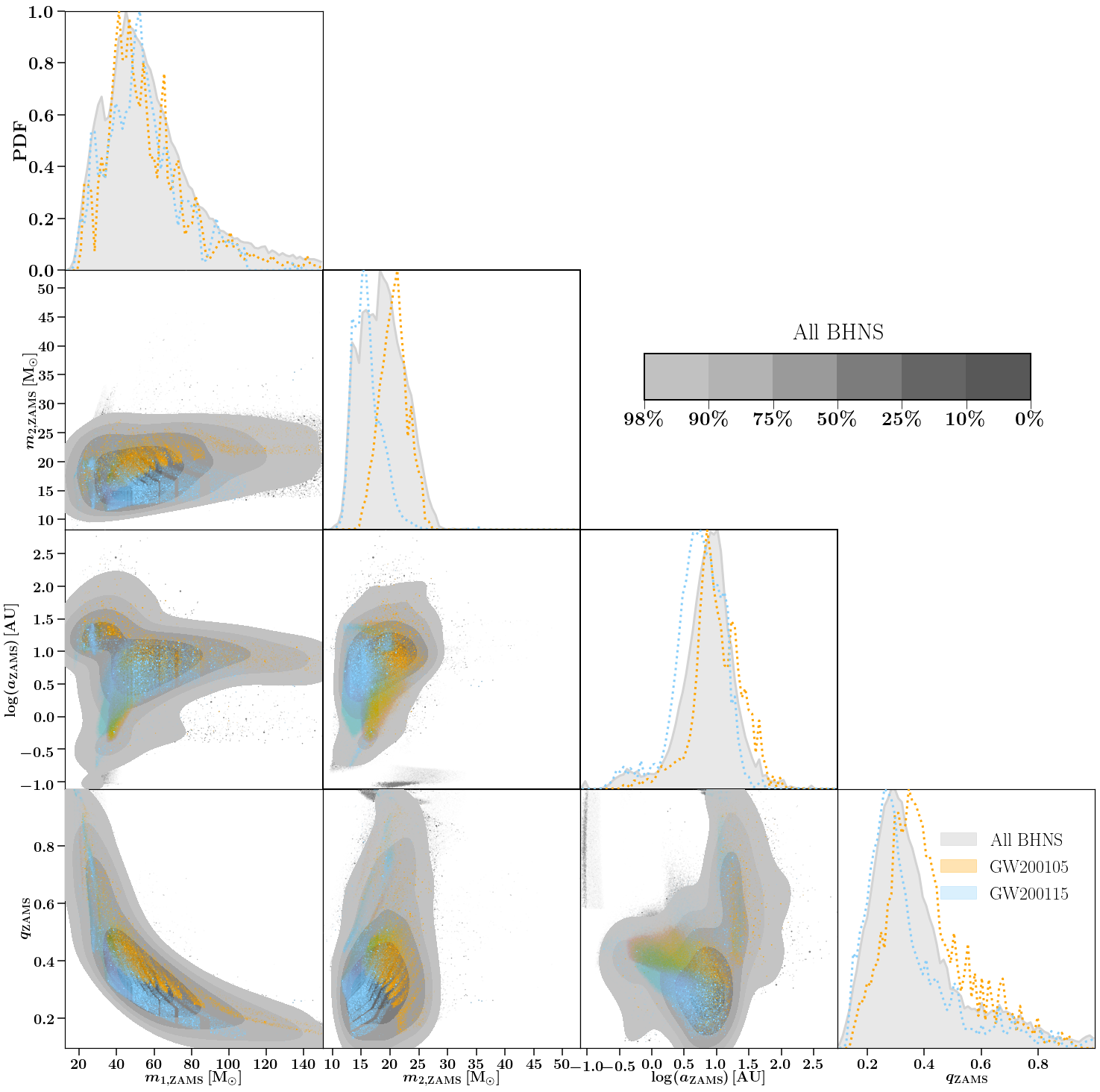}
    \caption{Same as Figure~\ref{fig:Triangle-final} but showing the  binary system properties at \ac{ZAMS} for the detectable \ac{BHNS} population of model \model. We show the primary mass, secondary mass, semimajor axis and mass ratio. We do not show inferred credible intervals from LIGO-Virgo in this figure as the ZAMS properties are not measurable through \acp{GW}. 
    }
    \label{fig:Triangle-ZAMS}
\end{figure*}

In Figure~\ref{fig:Triangle-ZAMS} we show the \ac{ZAMS} properties of the binary systems that successfully form detectable \ac{BHNS} mergers: primary mass (\monei), secondary mass (\mtwoi), semimajor axis (\ai), and mass ratio ($\qi \equiv \mtwoi/\monei$). In blue (orange) we show the ZAMS properties of binaries in our simulation that eventually form \ac{BHNS} matching the inferred credible intervals of \gwone (\gwzero). The distributions are weighted for the sensitivity of a \ac{GW}-detector network.  Several features can be seen that we describe below. 

First, we find that \gwone- and \gwzero-like \ac{GW} mergers form from binaries that have 1D distributions ($90$th percentiles) in the ranges $26 \lesssim \monei / \Msun \lesssim 112$,  $13 \lesssim \mtwoi / \Msun \lesssim 25$, $10^{0.04} \lesssim \ai / \AU \lesssim 10^{1.5}$, and $0.15 \lesssim \qi \lesssim  0.75$. From the histograms it can be seen that the initial conditions of the binaries that form \gwone- and \gwzero-like mergers are representative of the overall \ac{BHNS} forming population.  

Second, when comparing \gwone and \gwzero, we find that our model predicts that both systems formed from binaries with similar primary star masses. However, for the other ZAMS properties the model predicts that \gwzero-like \ac{BHNS} mergers form from binaries with slightly larger \mtwoi, \ai and \qi, compared to \gwone-like \ac{BHNS} mergers. The larger secondary masses for \gwzero are required to form the more massive \ac{NS} in this system. The larger secondary mass also causes the slight preference for larger \ai at \ac{ZAMS} as the increased secondary mass impacts the timing of mass transfer in several ways, including the time at which the primary will fill its Roche lobe, and the common-envelope phase later on (more/less shrinking due to a different envelope mass). As a result we find that \gwzero-like mergers form from slightly larger \ai compared to \gwone-like mergers.

Third, it can be seen that several of the distributions in Figure~\ref{fig:Triangle-ZAMS} show small gaps in ZAMS space that form \acp{BHNS} with combinations of \ac{BH} and \ac{NS} masses that do not match \gwone or \gwzero. These are mostly a consequence from small regions in \monei, \mtwoi and \qi that map to specific \ac{BH} masses in our stellar evolution prescriptions that do not match \gwone or \gwzero.

Finally, in the $\ai$--$\qi$ plane, we note a small population of \ac{BHNS} systems around $\log(\ai) \sim -1$ and $\qi \gtrsim 0.6$ that do not form \gwone- and \gwzero-like mergers. These are a small subset of \ac{BHNS} systems that form through an early mass transfer episode initiated by the primary star when it is still core-hydrogen burning (case A mass transfer). These systems are the main contributor to the small population of \acp{BHNS} in which the \ac{NS} forms first and with $\mbhf\gtrsim 10\Msun$ (see \citealt{Broekgaarden:2021} for further details).

%%%%%%%%%%%%%%%%%%%%%%%%%%%%%%%%%%%%%%%%%%%%%%%%
%%%%          DISCUSSION                 %%%%%%%
%%%%%%%%%%%%%%%%%%%%%%%%%%%%%%%%%%%%%%%%%%%%%%%%

\section{Discussion}
\label{sec:discussion}

\subsection{Predicted Merging BHNS Distribution Shapes for Models besides P112}
\label{sec:discussion-Distribution-shapes-of-models-besides-P112}

In \S\ref{sec:results-matching-the-GW-properties} we showed for model \model the predicted \ac{GW}-detectable  \ac{BHNS} distribution shapes; we now discuss how these are effected when considering the \Nmodels model realizations varying \SFRD and binary population synthesis assumptions.

First, changes in  \SFRD do not drastically impact the detectable \ac{BHNS} distribution shapes for  \mchirpf, \mtotf \mnsf, \mbhf, and \qf  as shown in \citet[][see Figure~14 and 15]{Broekgaarden:2021}. The predicted \ac{BHNS} merger rate density, on the other hand, is significantly impacted by the choice of \SFRD (with factors $\sim 10\times$;  Figure~\ref{fig:Rates-Intrinsic}). The \SFRD choice in particular also impacts the predicted \ac{BHBH} merger rate density, which puts the strongest constraints (out of all compact object merger flavors) on the matching \SFRD models in Figure~\ref{fig:Rates-Intrinsic} \PIItep.

Second, variations in binary stellar evolution assumptions do significantly impact the shape of the detectable \ac{BHNS} distributions for \mchirpf, \mtotf \mnsf, \mbhf and \qf as shown in \citet[][see Figure~14 and 15]{Broekgaarden:2021}.  Among the binary population synthesis models that match in rate (I, J, P and Q; \S\ref{sec:results-intrinsic-merger-rates}), model J ($\alpha_{\rm{CE}} = 10$) stands out as it predicts detectable \ac{BHNS} distributions that peak at low-mass events ($\mchirpf \lesssim 2\Msun$) compared to models I, P and Q (that peak near $\mchirpf \approx 3\Msun$).  The models with I ($\alpha_{\rm{CE}} = 2$) and Q ($\sigma = 30\kms$) have similar \ac{BHNS} distributions compared to model P112, with a small difference mainly in the tail of the mass distributions. We provide corner plots for the interested reader for these models in our  \href{https://github.com/FloorBroekgaarden/NSBH_GW200105_and_GW200115/tree/main/plottingCode/Fig_3_and_4/extra_figures}{Github repository}\footnote{\url{https://github.com/FloorBroekgaarden/NSBH_GW200105_and_GW200115/tree/main/plottingCode/Fig_3_and_4/extra_figures}} and refer the reader to \citet[][]{Broekgaarden:2021} for more details.  Future \ac{GW} observations might constrain between these models.

\subsection{Black Hole Spins and Neutron Star Tidal Disruption} 

\citet{Abbott:2021-first-NSBH} report the inferred $90\%$ credible interval for the primary spin magnitude $\chibh$  (i.e., spin of the \ac{BH}), of \gwone (\gwzero) to be  $\chibh = 0.33_{-0.29}^{+0.48}$  ($\chibh = 0.08_{-0.08}^{+0.22}$), while the spins of the \acp{NS} are unconstrained. Both reported \chibh values are consistent with zero. However,  for \gwone the authors report moderate support for negative \textit{effective} inspiral spin $\chi_{\rm{eff}} = -0.19_{-0.35}^{+0.23}$,
indicating a negatively aligned spin with respect to the orbital angular momentum axis. Theoretical studies of spins in \ac{BHNS} systems formed  through isolated binary evolution are still inconclusive. It has been argued that the black holes are expected to have $\chibh \approx 0$ due to  efficient angular momentum transport during the star's evolution  \citep[e.g.][]{FragosMcClintock:2015,FullerMa:2019}. Typically, no antialigned spins are expected \citep[but see, e.g., the discussion in ][]{Wysocki:2017,StegmannAntonini:2021}.  Studies including \citet[][]{Qin:2018} and \citet{Bavera:2020} argue that if the \ac{BH} is formed second, it can tidally spin up as a Wolf--Rayet (WR) star if the binary evolves through a tight \ac{BH}--WR phase. The same might be true for tight \ac{NS}--WR systems that can form \ac{BHNS} with a spun up \ac{BH} if the \ac{BH} forms after the \ac{NS} (e.g. if the system inverts its masses early in its evolution).  However, we find that none of the \gwone- and \gwzero-like \ac{BHNS} mergers in model \model do so, and hence  we predict $\chibh = 0$ for both events, consistent with the LIGO--Virgo inferred credible intervals.

Using the ejecta mass prescription from   \citet[][Equation~4]{Foucart:2018} and the \ac{BHNS} properties from model \model, we can crudely calculate whether our simulated \ac{BHNS} systems tidally disrupt the \ac{NS} outside the \ac{BH} innermost stable orbit and, if so, the amount of baryon mass outside the \ac{BH}. We find that when assuming $\chibh=0$ none of the \gwone- and \gwzero-like \ac{BHNS} systems have ejecta masses of $\gtrsim 10^{-6}$\Msun \citep[see][]{Abbott:2021-first-NSBH, Zhu:2021} for reasonable $\Rns = 11-13\km$.

\subsection{Other Formation Channels}

Previous predictions for \Rbhns from isolated binary evolution and alternative formation pathways have been made (see \citealt{MandelBroekgaardenReview:2021} for a review).  The various isolated binary evolution studies have predicted rates ranging from a  few tenths to $\sim 10^3$\Gpcyr, and a subset can match one of the LIGO--Virgo inferred \ac{BHNS} rates \citep[e.g.][]{Neijssel:2019,Belczynski:2020}. For the other formation channels, there are some studies that predict agreeable rates for formation from triples  \citep[e.g.][]{HamersThompson:2019}, formation in nuclear star clusters \citep[][but see also \citealt{PetrovichAntonini:2017, Hoang:2020}]{McKernan:2020}, dynamical formation in young star clusters \citep{Rastello:2020,Santoliquido:2020} and primordial formation \citep{Wang:2021}. On the other hand, much lower \ac{BHNS} rates ($\Rbhns\lesssim 10$\Gpcyr), which do not match the observed rate, are expected from binaries that evolve chemically homogeneously \citep{Marchant:2017}, from Population III stars \citep{Belczynski:2017popIII} and through dynamical formation in globular clusters \citep{Clausen:2013,ArcaSedda:2020,Hoang:2020,Ye:2019}. \ac{GW} observations of \ac{BHNS} might therefore provide a useful tool to distinguish between formation channels. We stress, however, that models should not only match the rates, but also the inferred mass and spin distributions of \ac{BHNS} mergers. This is particularly valuable as some of the formation channels predict \ac{BHNS} distributions with distinguishable features (e.g. a tail with larger \ac{BH} masses, $\mbhf\gtrsim 15-20\Msun$ in dynamical formation;  \citealt[][]{ArcaSedda:2020,Rastello:2020}) that could help constrain formation channels \citep[e.g.][]{Stevenson:2017spin}.

\subsection{Other Potential BHNS Merger Events}
Besides \gwone and \gwzero,  LVK reported four potential \ac{BHNS} candidates \citep{GWTC2,GWTC2point1}:
%
%%%%%
\begin{enumerate}
    \item GW190425 is most likely an NSNS merger, but a \ac{BHNS} origin cannot be ruled out.  If it is a \ac{BHNS} then $\mbhf = 2.0^{+0.6}_{-0.3}$\Msun and $\mchirpf = 1.44^{+0.02}_{-0.02}$\Msun are uncommon in our simulated \ac{BHNS} population (e.g., Figure~\ref{fig:Triangle-final} and \citealt[][]{Broekgaarden:2021}).
    \item GW190814 is most likely a BHBH merger, but a \ac{BHNS} origin cannot be ruled out. If so, it has $\mnsf =  2.59^{+0.008}_{-0.009}$\Msun. In \citet{Broekgaarden:2021} we noted that only our model K (which assumes a maximum NS mass of $3\Msun$) produces such heavy \ac{NS} masses, but that it does not form many GW190814-like \ac{BHNS} systems as GW190814's reported $\mchirpf =6.09^{+0.06}_{-0.06}$, $\mtotf = 25.8^{+1.0}_{-0.9}$ and $\mbhf=23.2^{+1.1}_{-1.0}$ are rare within the model \ac{BHNS} population. 
    \item GW190426$\_$152155 is a \ac{BHNS} candidate event, but with a marginal detection significance. If this event is real, it is inferred to have \ac{BHNS} properties very similar to \gwone \citep[see Figure~4 in][]{Abbott:2021-first-NSBH}
    %, namely $\mchirpf = 2.41^{+0.08}_{-0.08}$\Msun,  $\mtotf = 7.2^{+3.5}_{-1.5}$\Msun,  $\mbhf =  5.7^{+3.9}_{-2.3}$\Msun and $\mnsf = 1.5^{+0.8}_{-0.5}$\Msun. 
    We therefore predict it to be (similarly) common in our simulations.
    \item GW190917 is reported in the GWTC2.1 catalog, but the nature of its less massive component cannot be confirmed from the current data, and it was only classified as a BHBH event (i.e., $p_{\rm{BHNS}} = 0$) by the pipeline that detected it. If real, it might be a \ac{BHNS} with $\mchirpf = 3.7^{+0.2}_{-0.2}$, $\mtotf =11.4^{+3.0}_{-2.9}$, $\mbhf = 9.3^{+3.4}_{-4.4}$, $\mnsf = 2.1^{+1.5}_{-0.5}$ and $\qf =  0.23^{+0.52}_{-0.09}$. These properties are somewhat similar to \gwzero (although both the medians of \mbhf and \mnsf for GW190917 are slightly heavier), and we therefore predict it to be (similarly) common in our simulations.
\end{enumerate}
%%%%%%%%%%%%%%%%

\section{Conclusions}
\label{sec:conclusions}

In this Letter we studied the formation of the first two detected \ac{BHNS} systems (\gwone and \gwzero) in the isolated binary evolution channel using the \Nmodels binary population synthesis model realizations presented in \citet{ZenodoDCOBHNS:2021}. We investigate the predicted \Rbhns, as well as the \ac{BHNS} system properties (at merger and at ZAMS), and compare these with the data from LIGO--Virgo \citep{Abbott:2021-first-NSBH}. Our key findings are:

\begin{enumerate}
    \item We find that the majority of our \Nmodels model realizations can match one of the inferred credible intervals for \Rbhns from \citet{Abbott:2021-first-NSBH}. We further find that models with higher \ac{CE} efficiency ($\alpha_{\rm{CE}}\gtrsim 2$; models I and J) or moderate \ac{SN} natal kick velocities ($\sigma\lesssim 100\kms$; models P and Q) also match the inferred $90\%$ credible intervals for \Rbhbh and \Rnsns.
    
    \item Using model \model as an example, we find that the isolated binary evolution channel predicts a \ac{GW}-detectable \ac{BHNS} population that matches the observed properties (chirp mass, component masses, and mass ratios) of \gwone and \gwzero, although we expect a somewhat broader population than just \gwone- and \gwzero-like \ac{BHNS} systems. 
    
    \item We find that \gwone- and \gwzero-like \ac{BHNS} mergers form in model \model from binaries with ZAMS properties ($90\%$ percentiles) in the range $26 \lesssim \monei / \Msun \lesssim 112$,  $13 \lesssim \mtwoi / \Msun \lesssim 25$, $10^{0.04} \lesssim \ai / \AU \lesssim 10^{1.5}$ and $0.15 \lesssim \qi \lesssim 0.75$.  \gwone and \gwzero-like \ac{BHNS} systems have a similar range of primary star masses, but we expect \gwzero-like \ac{BHNS} mergers to form from binaries with slightly larger \mtwoi, \ai and \qi, compared to \gwone-like systems.
    
    \item We note that if \gwone and \gwzero were formed through isolated binary evolution, then we expect their \ac{BH} to have a spin of $\approx0$, their \ac{BH} to have formed first, and neither system to have produced an electromagnetic counterpart.
    
    \item We discuss the four other \ac{BHNS} candidates reported by LIGO--Virgo, and find that the properties of GW190425 and GW190814 do not match our predicted \ac{BHNS} population, making them instead more likely to be NSNS and BHBH mergers, respectively. On the other hand, the properties of the \ac{BHNS} candidates GW190426$\_$152155 and GW190917 do match our predicted \ac{BHNS} population, but were reported by LIGO-Virgo with low signal-to-noise ratios.
\end{enumerate}

We thus conclude that \gwone and \gwzero can be explained from formation through the isolated binary evolution channel, at least for some of the model realizations within our range of exploration.  With a rapidly increasing population of \ac{BHNS} systems expected in Observing Run 4 and beyond, it will be possible to carry out a more detailed comparison to model simulations, and to eventually determine the evolutionary histories of \ac{BHNS} systems.

%%%%%%%%%%%%%%%%%%%%%%%%%%%%%%%%%%%%%%%%%
\section*{Acknowledgements}

We thank Gus Beane, Debatri Chattopadhyay, Victoria DiTomasso, Griffin Hosseinzadeh,  Ilya Mandel, Noam Soker, Simon Stevenson, Alejandro Vigna-G\'{o}mez, Tom Wagg, Michael Zevin, and the members of TeamCOMPAS for useful discussions and help with the manuscript. We also thank the anonymous referee for helpful suggestions to this manuscript. This work was supported in part by the Prins Bernhard Cultuurfonds studiebeurs awarded to F.S.B. and by NSF and NASA grants awarded to E.B. 

 This research has made use of the publicly available binary population synthesis  simulations at doi: \href{https://doi.org/10.5281/zenodo.5178777}{10.5281/zenodo.5178777} (\citealt{ZenodoDCOBHNS:2021}, Broekgaarden et al. in preparation). These Simulations made use of the {COMPAS rapid binary population synthesis code} (\url{{http://github.com/TeamCOMPAS/COMPAS}}) including the STROOPWAFEL sampling algorithm \citep{Stevenson:2017,Barrett:2017,VignaGomez:2018,Broekgaarden:2019, Neijssel:2019}. 
 In addition, we used the posterior samples for \gwone and \gwzero provided by the Gravitational Wave Open Science Center (\url{https://www.gw-openscience.org/}), a service of LIGO Laboratory, the LIGO Scientific Collaboration and the Virgo Collaboration. LIGO Laboratory and Advanced LIGO are funded by the United States National Science Foundation (NSF) as well as the Science and Technology Facilities Council (STFC) of the United Kingdom, the Max-Planck-Society (MPS), and the State of Niedersachsen/Germany for support of the construction of Advanced LIGO and construction and operation of the GEO600 detector. Additional support for Advanced LIGO was provided by the Australian Research Council. Virgo is funded, through the European Gravitational Observatory (EGO), by the French Centre National de Recherche Scientifique (CNRS), the Italian Istituto Nazionale di Fisica Nucleare (INFN) and the Dutch Nikhef, with contributions by institutions from Belgium, Germany, Greece, Hungary, Ireland, Japan, Monaco, Poland, Portugal, Spain.
 This research has made use of NASA`s Astrophysics Data System Bibliographic Services.

\software{} Python version 3.6  \citep{CS-R9526},  Astropy \citep{astropy:2013,astropy:2018}, Matplotlib \citep{2007CSE.....9...90H},  {NumPy} \citep{2020NumPy-Array}, SciPy \citep{2020SciPy-NMeth}, \texttt{ipython$/$jupyter} \citep{2007CSE.....9c..21P, kluyver2016jupyter}, Pandas \citep{mckinney-proc-scipy-2010}, Seaborn \citep{waskom2020seaborn},  and  {hdf5}   \citep{collette_python_hdf5_2014}.

\section*{Data Availability}
All code to reproduce the figures and results in this work are publicly available at \url{https://github.com/FloorBroekgaarden/NSBH_GW200105_and_GW200115} corresponding to the Zenodo doi \href{https://zenodo.org/badge/latestdoi/395018301}{10.5281/zenodo.5541950}.


\bibliographystyle{aasjournal}
\bibliography{BHNS-MSSFR}{} % if your bibtex file is called example.bib

%%%%%%%%%%%%%%%%%%%%%%%%%%%%%%%%%%%%%%%%%%%%%%%%%%

%%%%%%%%%%%%%%%%% APPENDICES %%%%%%%%%%%%%%%%%%%%%

% \appendix
%%%%%%%%%%%%%%%%%%%%%%%%%%%%%%%%%%%%%%%%%%%
\listofchanges 

% Don't change these lines
% \bsp	% typesetting comment
% \label{lastpage}
\end{document}